\documentclass[sw]{iosart2x}

\usepackage[T1]{fontenc}
\usepackage{times}%
\usepackage{comment}
\usepackage{natbib}
\usepackage{amsmath}
\usepackage{dcolumn}
\usepackage{graphics}
\usepackage{graphicx}
\usepackage{url}
\usepackage{hyperref}
\usepackage{tabularx}
\usepackage{multirow}
\usepackage{comment}
\usepackage{tablefootnote}
\usepackage{caption}
\usepackage{subcaption}
\usepackage{stfloats}
\usepackage{capt-of}
\usepackage{xcolor}
\usepackage{afterpage}
\usepackage{multirow}
\usepackage{color}
\usepackage[np, autolanguage]{numprint}
\newcolumntype{d}[1]{D{.}{.}{#1}}
\firstpage{1} \lastpage{5} \volume{1} \pubyear{2018}

\begin{document}
\begin{frontmatter}                           % The preamble begins here.

%\pretitle{Pretitle}
\title{Empirical Methodology for Crowdsourcing Ground Truth}

\runningtitle{Empirical Methodology for Crowdsourcing Ground Truth}
%\subtitle{Subtitle}

%\review{Name Surname, University, Country}{Name Surname, University, Country}{Name Surname, University, Country}

\author[A]{\fnms{Anca} \snm{Dumitrache}},
\author[A]{\fnms{Oana} \snm{Inel}},
\author[C]{\fnms{Benjamin} \snm{Timmermans}},
\author[B]{\fnms{Carlos} \snm{Ortiz}}, 
\author[C]{\fnms{Robert-Jan} \snm{Sips}},
\author[A]{\fnms{Lora} \snm{Aroyo}} and
\author[D]{\fnms{Chris} \snm{Welty}}
\runningauthor{Dumitrache et al.}
\address[A]{Department of Computer Science, VU University, De Boelelaan 1081-1087, 1081 HV, Amsterdam,\\
E-mail: \{anca.dumitrache,oana.inel,lora.aroyo\}@vu.nl}
\address[B]{Netherlands eScience Center, Amsterdam, Netherlands,\\
E-mail: c.martinez@esciencecenter.nl}
\address[C]{CAS Benelux, IBM Netherlands,\\
E-mail: b.timmermans@nl.ibm.com, rhjsips@gmail.com}
\address[D]{Google Research, New York\\
E-mail: cawelty@gmail.com}

\begin{abstract}
The process of gathering ground truth data through human annotation is a major bottleneck in the use of information extraction methods for populating the Semantic Web. Crowdsourcing-based approaches are gaining popularity in the attempt to solve the issues related to volume of data and lack of annotators. Typically these practices use inter-annotator agreement as a measure of quality. However, in many domains, such as event detection, there is ambiguity in the data, as well as a multitude of perspectives of the information examples. We present an empirically derived methodology for efficiently gathering of ground truth data in a diverse set of use cases covering a variety of domains and annotation tasks. Central to our approach is the use of CrowdTruth metrics that capture inter-annotator disagreement. We show that measuring disagreement is essential for acquiring a high quality ground truth. We achieve this by comparing the quality of the data aggregated with CrowdTruth metrics with majority vote, over a set of diverse crowdsourcing tasks: {\it Medical Relation Extraction}, {\it Twitter Event Identification}, {\it News Event Extraction} and {\it Sound Interpretation}. We also show that an increased number of crowd workers leads to growth and stabilization in the quality of annotations, going against the usual practice of employing a small number of annotators.
\end{abstract}

\end{frontmatter}

\section{Introduction}
\label{sec:introduction}

Knowledge base curation, or the task of populating knowledge bases, is one of the main research challenges of crowdsourcing the Semantic Web~\cite{sarasua2015crowdsourcing}. Knowledge base curation can be done either manually, by asking annotators to populate the knowledge graph by manually extracting triples from unstructured data, or automatically by using information extraction methods that are trained and evaluated on ground truth collected from human annotators. In both cases, the process of gathering the human annotations is the a bottleneck in the entire knowledge base population process. The traditional approach to gathering human annotation is to employ experts to perform annotation tasks~\cite{welty2012query}, which is a costly and time consuming process. Additionally, in order to prevent high disagreement among expert annotators, strict annotation guidelines are designed for the experts to follow. On the one hand, creating such guidelines is a lengthy and tedious process, and on the other hand, the annotation task becomes rigid and not reproducible across domains. And, as a result, the entire process needs to be repeated over and over again in every domain and task. Finally, expert annotators are not always available for specific tasks such as open domain question-answering or news events, while many annotation tasks can require multiple interpretations that a single annotator cannot provide~\cite{aroyo2012harnessing}.

As a solution to those problems, crowdsourcing has become a mainstream approach. It has proved to provide good results in multiple domains: annotating cultural heritage prints~\cite{oosterman2014crowdsourcing}, medical relation annotation~\cite{aroyo2013measuring}, ontology evaluation~\cite{noy2013mechanical}. Following the central feature of volunteer-based crowdsourcing introduced by~\cite{von2009human} that majority voting and high inter-annotator agreement~\cite{Carletta1996} can ensure truthfulness of resulting annotations, most of those approaches are assessing the quality of their crowdsourced data based on the hypothesis~\cite{nowak2010reliable} that there is only one right answer to each question.

However, this assumption often creates issues in practice. Recent work in collecting annotations for text~\cite{poesio2005reliability,chang2016linguistic}, sounds~\cite{doi:10.1080/09298215.2016.1200631} and images~\cite{schaekermann2016,cheplygina2018crowd} found that disagreement between annotators is not just a result of poor quality work, and can actually be an indicator for other properties of the data, such as ambiguity and uncertainty~\cite{aroyo2018aimag}.

Previous experiments we performed~\cite{aroyo2013crowd} also identified issues with the assumption of the one truth: inter-annotator disagreement is usually never captured, either because the number of annotators is too small to capture the full diversity of opinion, or because the crowd data is aggregated with metrics that enforce consensus, such as majority vote.  These practices create artificial data that is neither general nor reflects the ambiguity inherent in the data.

To address these issues, we proposed the {\bf CrowdTruth} methodology for crowdsourcing human annotation by harnessing inter-annotator disagreement, i.e representing the diversity of human interpretations in the ground truth. This is a novel approach for crowdsourcing human annotation that, instead of enforcing agreement between annotators, captures the ambiguity inherent in semantic annotation through the use of ambiguity-aware metrics for aggregating crowdsourcing responses.  Based on this principle, we have implemented the CrowdTruth methodology as part of a framework~\cite{inel2014crowdtruth} for machine-human computation, that first introduced the ambiguity-aware metrics and built a pipeline to process crowdsourcing data with these metrics.

In this paper, we extend the definition of our ambiguity-aware methodology (CrowdTruth version 1.0~\cite{inel2014crowdtruth}) to work both with crowdsourcing tasks that are {\it closed}, i.e. the annotations that can occur in the data are already known, and the workers are asked to validate their existence (e.g. given a news event, decide whether it is expressed in a tweet), and tasks that are {\it open}, i.e. the annotation space is not known, and workers can freely select all the choices that apply (e.g. given a news piece, select all events that appear in the text). The code for the extended CrowdTruth version 1.1 methodology and metrics is available at: \url{https://git.io/fA3Mq}.

We investigate tasks of text and sound annotation, in both domains that typically require expertise from annotators (e.g. medical) and those that don't (open domain).  In particular, we look at four crowdsourcing tasks: {\it Medical Relation Extraction}, {\it Twitter Event Identification}, {\it News Event Extraction} and {\it Sound Interpretation}.  The aim is to investigate the role of inter-annotator disagreement as part of the crowdsourcing system by applying the CrowdTruth methodology to collect data over a set of diverse use cases. 

Through the use of CrowdTruth aggregation metrics, the interpretations collected from the crowd are transformed into explicit semantics for the various tasks presented in this paper -- i.e. relations expressed in sentences, topics / events expressed in tweets and news articles, words describing sounds -- thus enabling knowledge base curation for these specific tasks.  Furthermore, we prove that capturing disagreement is essential for acquiring high quality semantics.  We achieve this by comparing the quality of the data aggregated with CrowdTruth metrics with majority vote, a method which enforces consensus among annotators.  By applying our analysis over a set of diverse tasks we show that, even though ambiguity manifests differently depending on the task (e.g. each task has an optimal number of workers necessary to capture the full spectrum of opinions), our theory of inter-annotator disagreement as a property of ambiguity is generalizable for any semantic annotation crowdsourcing task.

The paper makes the following contributions:

\begin{enumerate}

\item {\bf comparative analysis of crowdsourcing aggregation methods:} we compare the performance of {\it ambiguity-aware metrics} and {\it consensus - enforcing metrics} over a diverse set of crowdsourcing tasks (Sections \ref{sec:results}, \ref{sec:discussion});

\item {\bf stability of crowd results:} we show in several crowdsourcing tasks that {\it an increased number of crowd workers leads to growth and stabilization in the quality of annotations}, going against the usual practice of employing a small number of annotators (Sections \ref{sec:results}, \ref{sec:discussion});

\item {\bf measuring quality in open-ended tasks:} we present an extension to the CrowdTruth methodology that allows the ambiguity-aware metrics to deal {\it both with open-ended and closed tasks} (Sections \ref{sec:methodology}, \ref{sec:experimental_setup}), as opposed to the initial version of the CrowdTruth metrics which only processed closed tasks;

\item {\bf semantics of ambiguity:} applying the CrowdTruth methodology we collect richer data that allows to reason about ambiguity of content (in all modality formats, e.g. images, videos and sounds), which is intrinsically relevant to the Semantic Web community.

\end{enumerate}

\section{CrowdTruth Methodology}
\label{sec:methodology}

In this section, we describe the CrowdTruth {\it methodology} version 1.1, for aggregating crowdsourcing data, which offers methods to aggregate both closed an open-ended tasks. Version 1.1 presented in this paper is a generalization of the initial version 1.0 of CrowdTruth~\cite{inel2014crowdtruth}.

In Section~\ref{sec:results} we use a number of annotation tasks in different domains  to illustrate its use and gather experimental data to prove the main claim of this research - CrowdTruth methodology provides a viable alternative to traditional consensus-based majority vote crowdsourcing and expert-based ground truth collection. The elements of the CrowdTruth methodology are:
\begin{itemize}
\item annotation modeling with the \emph{triangle of disagreement};
\item quality \emph{metrics} for media units (input data), annotations and crowd workers;
\item identification of workers with low quality annotations.
\end{itemize}

Each of these elements is applicable across a variety of domains, content modalities, \emph{e.g.}, text, sounds, images and videos and annotation tasks, \emph{e.g.}, closed and open-ended annotations. The following sub-sections briefly introduce the overview of the methodology elements.

\subsection{CrowdTruth quality metrics}
\label{subsec:metrics}

\begin{figure}[!hpt]
 	\centering
 		\includegraphics[width=0.8\linewidth]{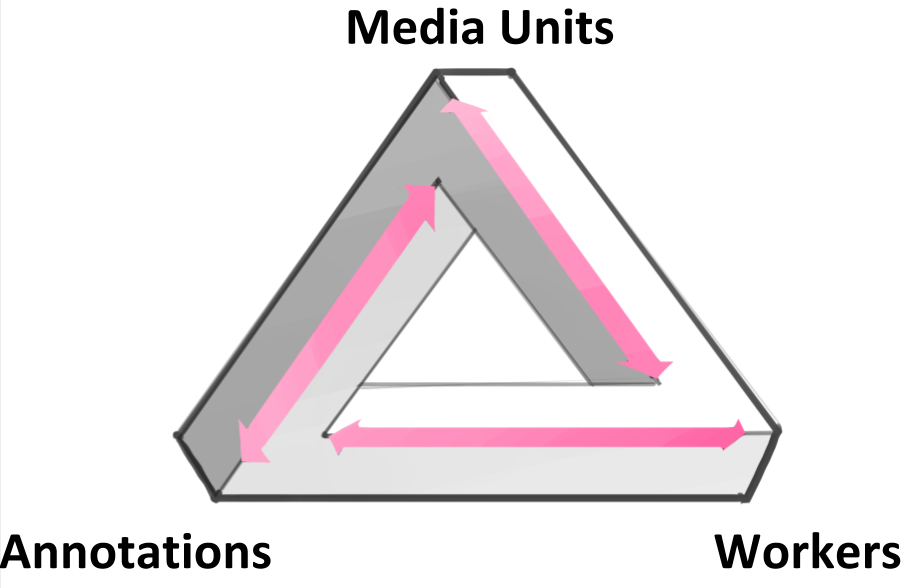}
 	\caption{Triangle of Disagreement}
 	\label{fig:triangle_of_reference}
 \end{figure}

Measuring quality in CrowdTruth is done with  the triangle of disagreement model (based on the triangle reference \cite{knowlton1966definition}), which links together media units, workers, and annotations, as seen in Fig.\ref{fig:triangle_of_reference}. It allows us to assess the quality of each worker, the clarity of each media unit, and the ambiguity, similarity and frequency of each annotation. This model makes it possible to express how the ambiguity in any of the corners disseminates and influences the other components of the triangle. For example, an unclear sentence or an ambiguous annotation scheme would cause more disagreement between workers \cite{aroyo2014threesides}, and thus, both need to be accounted for when measuring the quality of the workers. 

The CrowdTruth quality metrics~\cite{aroyo2014threesides} are designed to capture inter-annotator disagreement in crowdsourcing. The metrics were introduced for {\it closed tasks}, i.e. multiple choice tasks, where the annotation set is known before running the crowdsourcing task. In this paper, we present an extended version of these metrics (version 1.1), that can be used for both {\it closed tasks} as well as {\it open-ended tasks} (i.e. the annotation set is not known beforehand, and the workers can freely select all the choices that apply). The code for the CrowdTruth version 1.1 metrics is available at: \url{https://git.io/fA3Mq}.

The quality of the crowdsourced data is measured using a \textbf{vector space representation} of the crowd annotations. For \emph{closed tasks}, the annotation vector contains the given answer options in the task template, which the crowd can choose from. For example, the template of a \emph{closed task} can be composed of a multiple choice question, which appears as a list checkboxes or radio buttons, thus, having a finite list of options to choose from.

While for closed tasks the number of elements in the annotation vector is known in advance, for \emph{open-ended tasks} the number of elements in the annotation vector can only be determined when all the judgments for a media unit have been gathered. An example of such a task can be highlighting words or word phrases in a sentence, or as an input text field where the workers can introduce keywords. In this case the answer space is composed of all the unique keywords from all the workers that solved that media unit. As a consequence, all the media units in a closed task have the same answer space, while for open-ended tasks the answer space is different across all the media units.

Although the answer space for open-ended tasks is not known from the beginning, it is still possible to deduce a finite answer space. To achieve this, we added an {\it answer space dimensionality reduction step} to the methodology for open-ended tasks. Additional goals of this step are to reduce redundancy in the answer space through similarity clustering (e.g. by making sure that synonymous words do not count as disagreement between annotators), and to keep the vector space representation small enough so that the CrowdTruth quality metrics still produce meaningful values. The method for performing dimensionality reduction is dependent on the annotation task itself. % Different methods are needed when dealing with, for instance, tasks that give free text input to the crowd vs. tasks that ask the crowd to combine various)

In the annotation vector, each answer option is a boolean value, showing whether the worker annotated that answer or not. This allows the annotations of each worker on a given media unit to be aggregated, resulting in a \textbf{media unit vector} that represents for each option how often it was annotated.

Three core \textbf{worker metrics} are defined to differentiate between low-quality and high-quality workers. \emph{Worker-Worker Agreement} ($wwa$) measures the pairwise agreement between two workers across all media units they annotated in common - indicating how close a worker performs compared to workers solving the same task. \emph{Worker-Media Unit Agreement} ($wma$) measures the similarity between the annotations of a worker and the aggregated annotations of the rest of the workers. The average of this metric across all the media units solved gives a measure of how much a worker disagrees with the crowd in the context of all media units. \emph{Average annotations per media unit} ($na$) measures for each worker the total number of annotations they chose per media unit, averaged across all media units they annotated. Since in many tasks workers can choose all the possible annotations, a low quality worker can appear to agree more with the rest of the workers by repeatedly choosing multiple annotations, thus increasing the chance of overlap.

\begin{table*}
	\caption {Consider an open-ended sound annotation task where 10 workers have to describe a given sound with keywords. The media unit for this task is a sound, the annotation set contains all the keywords workers provide for a sound. The table shows the media unit metrics, as well as the majority vote score for the media unit.}
    \label{tab:example_annotation}
	\begin{tabular}{ | r | c c c c c |}
    \hline
    {\bf worker annotations} & {\it dog barking} & {\it walking} & {\it animal} & {\it echo} & {\it loud} \\
    {\bf media unit vector} & 3 & 2 & 5 & 1 & 1 \\
    {\bf media unit -- annotation score} &  0.47 & 0.31 & 0.79 & 0.15 & 0.15 \\ 
    {\bf majority vote} &  0 & 0 & 1 & 0 & 0 \\ \hline
    \end{tabular}
\end{table*}

Two \textbf{media unit metrics} are defined to assess the quality of each unit. In this paper, we focus on the \emph{Media Unit-Annotation Score} -- the core CrowdTruth metric, used to measure the clarity with which the media unit expresses a given annotation. This metric is computed for each media unit and each possible annotation as the cosine between the media unit vector and the unit vector for each possible annotation.  This metric is used in evaluating the quality of the CrowdTruth annotations.

\subsection{Spam Removal}
\label{subsec:spam_removal}

After collecting the crowd annotations, but before the evaluation of the data, we perform spam removal.  The purpose of this step is to identify the adversarial and low quality workers -- e.g. those workers that always pick the same annotations, regardless of the unit. Once identified, the spam workers are removed from the dataset, and their annotations are not used in the evaluation.  The methodology for spam removal is based on our previous work in~\cite{soberon2013}, extended in this paper to work also for open-ended tasks. 

We identify the low quality workers by applying the core CrowdTruth worker metrics, the worker-worker agreement ($wwa$), worker-media unit agreement ($wma$) and the average number of annotations ($na$) submitted by a worker for one sentence. The first two metrics are used to model the extent to which a given worker agrees with the other annotators. The purpose is not to penalize disagreement with the majority, but rather to identify outliers, \emph{i.e.}, workers that are in constant disagreement. For \emph{closed tasks} where the semantics of the annotations in the answer space could rarely overlap, it is unlikely that a large number of possible annotations will occur for the same media unit. Therefore, the number of annotations per sentence can also indicate spam behavior. 

In \emph{open-ended tasks} we apply the same approach. However, we need to acknowledge the fact that open-ended tasks are more prone to disagreement due to the large answer space and thus, the overall agreement between the workers can occur with lower values. Thus, we do not have predefined values for identifying the low-quality workers, but for every task or job we use the following main heuristic: given worker $w$, if the agreement $wwa(w)$, $wsa(w)$ and optionally, annotations per sentence $na(w)$, parameters do not fall within the standard deviation for the task, then worker $w$ is marked as a spammer. To confirm the validity of this metrics we also perform manual evaluation based on sampling of the results.

Based on the specificity of each task, closed or open-ended, the effort required to pick different annotations might vary. For instance, when no good annotation exists in the media unit, the time to complete the annotation is considerably reduced. This can bias the workers towards selecting the option that requires the least work. In order to prevent this, we introduce {\it in-task effort consistency checks}. Such annotations do not count towards building the ground truth, and are used to reduce the bias from picking the quickest option. For instance, when stating that no annotation is possible in the media unit, the workers also have to write an explanation in a text box for why no annotation were provided.

%ANCA you might want to say here that we try to make all answer options equally effort consuming, so that there is no stimulous for only giving the easy answers, and thus we ensure higher quality of the results.

\begin{table*}[!htp]
	\caption {Crowdsourcing Task Details}
	\begin{tabular}{ | r | c c l |}
    \hline
    {\bf Task} & {\bf Type} & {\bf Media Unit} & {\bf Annotations} \\
    \hline \hline
    \multirow{3}{*}{Medical Relation Extraction} & \multirow{3}{*}{closed} & \multirow{3}{*}{sentence} & medical relations: {\it cause}, {\it treat}, {\it prevent}, {\it symptom}, {\it diagnose},  \\
    & & & {\it side effect}, {\it location} {\it manifestation}, {\it contraindicate}, \\
    & & & {\it is a}, {\it part of}, {\it associated with}, {\it other}, {\it none} \\ \hline
    \multirow{5}{*}{Twitter Event Identification} & \multirow{5}{*}{closed} & \multirow{5}{*}{tweet} & tweet events: {\it Davos world economic forum 2014}, {\it FIFA World Cup 2014},  \\
    & & & {\it Islands disputed between China and Japan}, {\it 2014 anti-China protests in Vietnam}, \\
    & & & {\it Korean MV Sewol ferry ship sinking}, {\it Japan whaling and dolphin hunting},  \\
    & & & {\it Disappearance of Malaysia Airlines flight 370}, {\it Ukraine crisis 2014}, \\ 
    & & & {\it none of the above} \\ \hline
    News Event Extraction & open-ended & sentence & words in the sentence \\ \hline
    Sound Interpretation & open-ended & sound & tags describing sound \\ \hline
    \end{tabular}
    \label{tab:crowd_data}
\end{table*}

\begin{table*}[!htp]
	\caption {Crowdsourcing Task Data}
	\begin{tabular}{ | r | c c c c c | }
    \hline
    {\bf Task} & {\bf Source} & {\bf Expert annotation} & {\bf Media Units} & {\bf Workers / Unit} & {\bf Cost / Judgment} \\
    \hline \hline
    Medical Relation Extraction & PubMed article abstracts & yes & 975 & 15 & \$0.05  \\
    Twitter Event Identification & Twitter (2014) & no & \np{3019} & 7 & \$0.02 \\
    News Event Extraction & TimeBank & yes & \np{200} & 15 & \$0.02 \\
    Sound Interpretation & Freesound.org & yes  & \np{284} & 10 & \$0.01 \\ \hline
    \end{tabular}
    \label{tab:crowd_tasks}
\end{table*}

\section{Experimental Setup}
\label{sec:experimental_setup}

The aim of the crowdsourcing experiments described and analyzed in this paper is to show that the CrowdTruth ambiguity-aware crowdsourcing approach produces data with a higher quality than the traditional majority vote where consensus among annotators is enforced. In order to show this, we perform an experiment over a set of four diverse crowdsourcing tasks: 
\begin{itemize}
    \item two closed tasks, i.e. {\it Medical Relation Extraction}, {\it Twitter Event Identification},
    \item two open-ended tasks, i.e. {\it News Event Extraction} and {\it Sound Interpretation}.
\end{itemize}
These tasks were picked from diverse domains (medical, sound, open), to aid in the generalization of our results.  To evaluate the quality of the crowdsourcing data, we constructed a trusted judgments set by combining expert and crowd annotations. The rest of this section describes the details of the crowdsourcing tasks, trusted judgments acquisition process, as well as the evaluation methodology we employed.

\subsection{Crowdsourcing Overview}
\label{subsec:detasets}

\begin{figure*}[!tb]
\centering
\caption{Templates of the Crowdsourcing Tasks}
\label{fig:taskdesigns}
\begin{subfigure}{.45\textwidth}
\includegraphics[width=\linewidth]{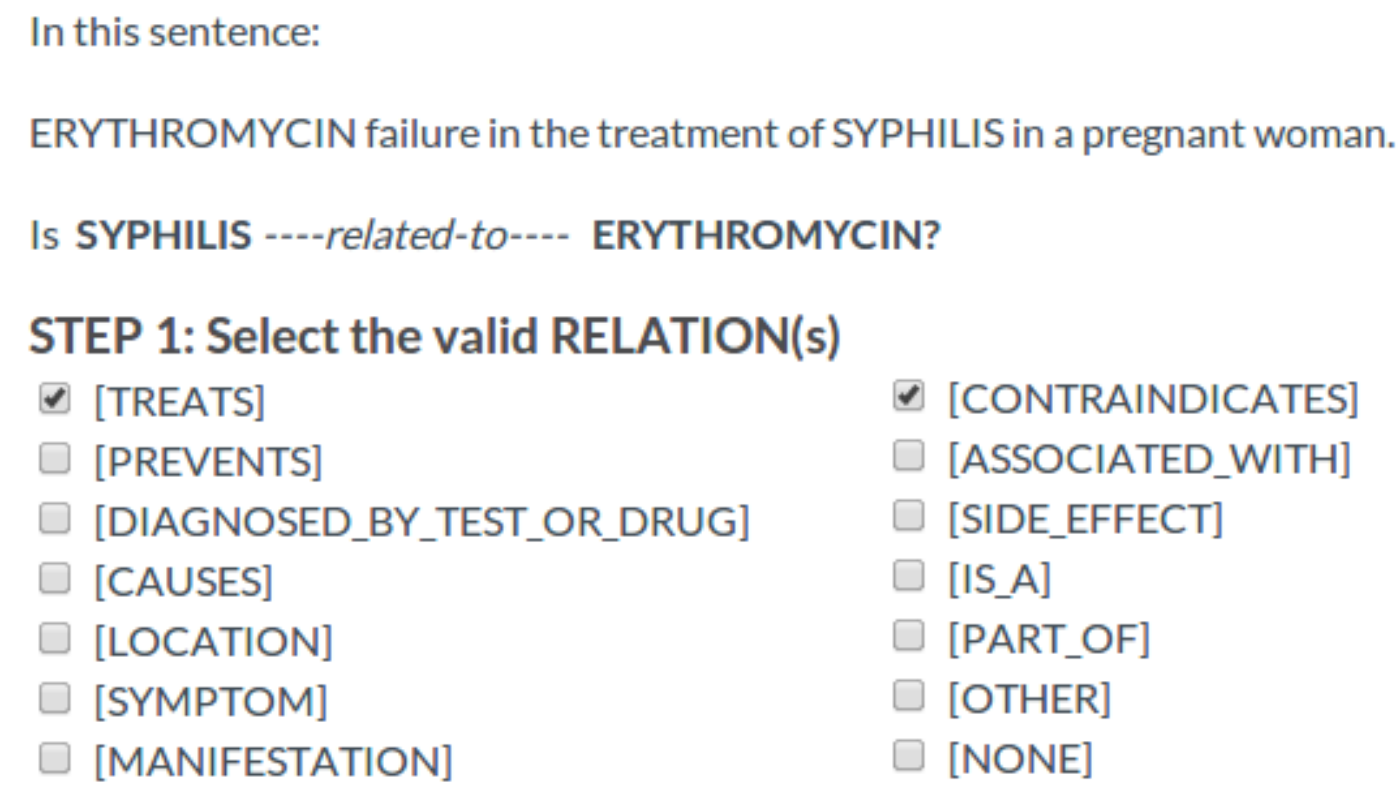}
\caption{Medical Relation Extraction}
\label{fig:screenshot_medical}
\end{subfigure}%
\begin{subfigure}{.45\textwidth}
\includegraphics[width=\linewidth]{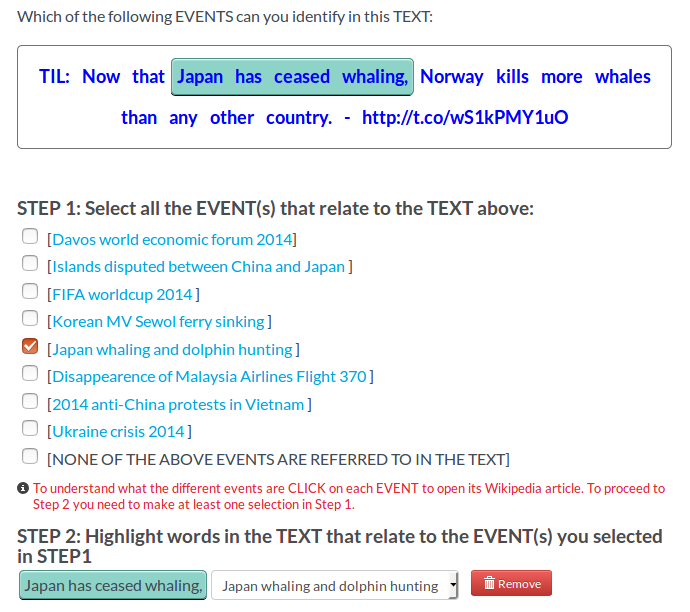}
\caption{Twitter Event Identification}
\label{fig:screenshot_tweets}
\end{subfigure}
\begin{subfigure}{.45\textwidth}
\includegraphics[width=\linewidth]{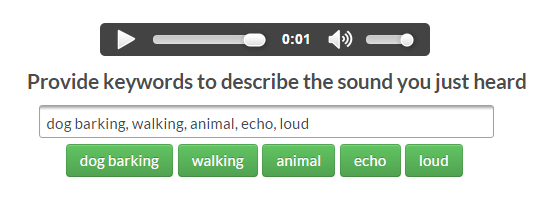}
\caption{Sound Interpretation}
\label{fig:screenshot_sounds}
\end{subfigure}%
\begin{subfigure}{.45\textwidth}
\includegraphics[width=\linewidth]{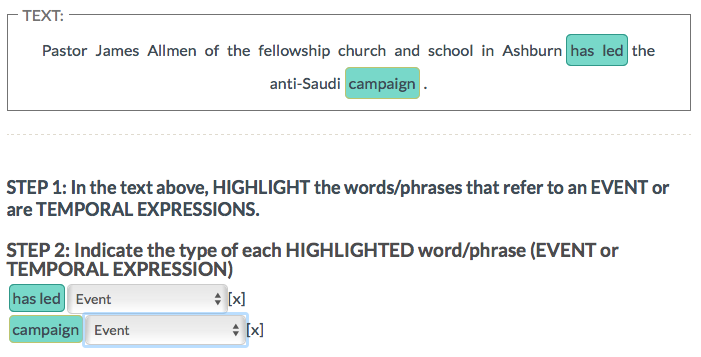}
\caption{News Event Extraction}
\label{fig:screenshot_news}
\end{subfigure}
\end{figure*}

Tables \ref{tab:crowd_data} and \ref{tab:crowd_tasks} present an overview of the crowdsourcing tasks, as well as the datasets used. The results of the crowdsourcing tasks were processed with the use of CrowdTruth metrics (Sec.~\ref{subsec:metrics}), and we removed consistently low quality workers based on the spam removal procedure (Sec~\ref{subsec:spam_removal}). The tasks were implemented and ran on Figure Eight\footnote{\url{https://figure-eight.com/}} (formerly known as CrowdFlower). The templates are available on the CrowdTruth platform\footnote{tasks marked with $*$: \url{https://github.com/CrowdTruth/CrowdTruth/wiki/Templates}}. 

The payment per judgment was determined through a series of pilot runs of the tasks where we started with a \$0.01 cost per judgment, and then gradually increased the payment until a majority of Figure Eight workers rated our tasks as having fair payments. As a result, we were able to get a constant stream of workers to participate in the tasks. The values shown in Table~\ref{tab:crowd_tasks} show the final cost per judgment we reached after the pilot runs. Since crowd pay has a complex effect on the quality of the annotation~\cite{mao2013volunteering}, and in order to remove confounding factors, judgments collected with costs lower than those in Table~\ref{tab:crowd_tasks} were left out of this evaluation. In total, it took two months to perform the pilot runs and then collect the judgments for all of the tasks.

The number of workers per media unit was determined experimentally with the goal of capturing all possible results from the crowd and stabilizing the quality of the annotations; this process is explained at length further on in Section~\ref{sec:results}, with the results of the experiment shown in Figure~\ref{fig:f1_workers}.

The {\bf Medical Relation Extraction dataset} consists of 975 sentences extracted from PubMed\footnote{\url{http://www.ncbi.nlm.nih.gov/pubmed}} article abstracts. The sentences were collected using distant supervision~\cite{mintz2009distant}, a method that picks positive sentences from a corpus based on whether known arguments of the seed relation appear together in the sentence (\emph{e.g.}, the $treat$ relation occurs between the terms $antibiotics$ and $typhus$, so find all sentences containing both and repeat this for all pairs of arguments that hold). The MetaMap parser~\cite{aronson2001effective} was used to extract medical terms from the corpus and the UMLS vocabulary~\cite{bodenreider2004unified} was used for mapping terms to categories, and relations to term types. The intuition of distant supervision is that since we know the terms are related, and they are in the same sentence, it is more likely that the sentence expresses a relation between them (than just any random sentence). We started with a set of 8 UMLS relations important for clinical decision making~\cite{P14-1078}, that became the seed in distant supervision, but this paper only discusses results for the relations $cause$ and $treat$, as these were the only relations for which we could also collect expert annotations. The expert judgment collection is detailed in Section~\ref{subsec:expert_data}.

The \emph{medical relation extraction task} (see Figure \ref{fig:screenshot_medical}) is a {\it closed task}. The crowd is given a medical sentence with the two highlighted terms collected with distant supervision, and is then asked to select from a list all relations that are expressed between the two terms in the sentence. The relation list contains eight UMLS\footnote{https://www.nlm.nih.gov/research/umls/} relations, as well as {\it is a}, {\it part of}, {\it associated with}, {\it other}, {\it none} relations, added to make the choice list complete. Multiple choices are allowed in this task. To reduce the bias of selecting $none$, we also added an in-task effort consistency check by asking workers to explain in a text box why no relation is possible between the terms. The task results are processed into an annotation vector containing a component for each of the relations. A detailed description of the crowdsourcing data collection is given in \cite{DBLP:journals/corr/DumitracheAW17}.

The {\bf Twitter Event Identification dataset} consists of 3,019 English tweets from 2014, crawled from Twitter. The tweets are selected as been relevant to eight events, such as, ``Japan whale hunt'', ``China Vietnam relation'' among other controversial events. The dataset was created by querying a Twitter dataset from 2014 with relevant phrases for each of the eight events, \emph{e.g.}, ``Whaling Hunting'', ``Anti-Chinese in Vietnam''. The \emph{Twitter event identification task} (see Figure \ref{fig:screenshot_tweets}) is a {\it closed task}. The crowd is asked to choose for each tweet the relevant events out of the list of eight, as well as to highlight for each of the relevant events the event mentions in the tweet. The crowd could also pick that none of the events was present in the tweet. Multiple choices of events were permitted. Since tweets and tweet annotations typically are not done by experts, we did not collect expert data for this task. To reduce the bias of selecting no event, we also added an in-task effort consistency check by asking workers to explain in a text box why none of the events is present in the tweet. The task results are processed into an annotation vector containing a component for each of the events.

The {\bf News Event Extraction dataset} consists of 200 randomly selected English sentences from the English TimeBank corpora~\cite{pustejovsky2003timebank}, which were also presented in \cite{CASELLI16.966}. The \emph{news event extraction} (see Figure \ref{fig:screenshot_news}) is an {\it open-ended task}. The crowd receives an English sentence, and is asked to highlight words or word phrases (multiple words) that describe an event or a time expression. For each sentence, the crowd is allowed to highlight a maximum of 30 event expressions or time expressions. For the purpose of this research we only focus on evaluating the extraction of event expressions. We define an \emph{event} as something that happened, is happening, will or happen. On this dataset we employed expert annotators as described in Section \ref{subsec:expert_data}. To reduce the bias of selecting fewer events than actually expressed in the task, we implemented an in-task effort consistency check by asking workers that annotated 3 events or less to explain in a text box why no other events are expressed in the sentence. As part of the {\it answer set dimensionality reduction step}, we removed the stop words from the sentence (we consider that the stop words are not meaningful for our analysis and they could add unsubstantial disagreement), and split the expressions collected from the crowd into words.  The annotation vector is composed of the words in the sentence, where a word is selected in the worker vector if it appears in at least one of the expressions identified by the worker.

The {\bf Sound Interpretation dataset} consists of 284 unique sounds sampled from the Freesound\footnote{\url{https://www.freesound.org/}} online database. All these recordings and their metadata are freely accessible through the Freesound API\footnote{\url{https://www.freesound.org/docs/api/}}. We focused on SoundFX sounds, \emph{i.e.}, sound effects category, as classified by \cite{font2014audio}. The \emph{Sound interpretation task} (see Figure \ref{fig:screenshot_sounds}) is an {\it open-ended task}, where the crowd is asked to listen to three sounds and provide for each sound a comma separated list of keywords that best describe what they heard. For each sound, any number of answers is possible. In the {\it answer set dimensionality reduction step}, the annotated keywords were clustered syntacticly using spell checking and stemming, and semantically using a word2vec model~\cite{mikolov2013distributed} pre-trained on the Google News corpus.  The annotation vector contains a component for each of the keywords used to describe the sound, after clustering. A detailed description of the crowdsourcing data collection and processing is given in \cite{miltenburg2016}. For this dataset we also collected expert annotations from the sound creators as described in Section \ref{subsec:expert_data}.

\subsection{Evaluation Methodology}
\label{subsec:evaluation}

The purpose of the evaluation is to determine the quality of the annotations generated with CrowdTruth ambiguity-aware aggregating metrics. To this end, we label each media unit and annotation pair with its media unit-annotation score (see Section \ref{subsec:metrics}), and compare it with three other methods for labeling the data, as described below:

\begin{itemize}
\item \textbf{Majority vote}:  Each media unit-annotation pair receives either a positive or a negative label, according to the decision of the majority of crowd workers. For each annotation performed by a crowd worker over a given media unit, we calculate the ratio of workers that have selected this annotation over the total number of workers that have annotated the unit, and assess whether it is greater or equal to 0.5. This allows for multiple annotations to be picked for one media unit. For some units, however, none of the annotations were picked by half or more of the workers. This is especially the case for open-ended tasks, such as sound interpretation, where workers put in a large number of annotations, and agreement is seldom. In these situations, we picked the annotations that were selected by the most workers (even if they do not constitute more than half). Judgments from workers labeled as spammers were not employed in the aggregation. An example of the majority vote aggregation is shown in Table~\ref{tab:example_annotation}.

\item \textbf{Single}:  Each media unit-annotation pair receives either a positive or a negative label, according to the decision of a single crowd worker. For every media unit, this score was randomly sampled
from the set of workers annotating it. Judgments from workers labeled as spammers were not employed. While a single annotator is not used as often as the majority vote in traditional crowdsourcing, we use this dataset as a baseline for the crowd, to show that having more annotators generates better quality data.

\item \textbf{Expert}: Each media unit-annotation pair receives either a positive or a negative label, according to the expert decision. The details of how expert data was collected for each tasks are discussed in Section \ref{subsec:expert_data}.
\end{itemize}

The \emph{evaluation of the quality of the CrowdTruth method} was done by computing the micro-F1 score over each task. The micro-F1 score was used in order to treat each case equally, without giving advantage to annotations that appear less frequently in our datasets. Using the trusted judgments collected according to Section~\ref{subsec:expert_data}, we evaluate each media unit -- annotation pair as either a true positive, false positive etc. We compute the value of the micro-F1 score using the following formulas for the micro precision (Equation \ref{eq:precision}) and micro recall (Equation \ref{eq:recall}):

\begin{equation}
P_{micro} = \frac{\sum_{i=1}^{n}{TP_i}}{\sum_{i=1}^{n}{TP_i} + \sum_{i=1}^{n}{FP_i}}
\label{eq:precision}
\end{equation}

\begin{equation}
R_{micro} = \frac{\sum_{i=1}^{n}{TP_i}}{\sum_{i=1}^{n}{TP_i} + \sum_{i=1}^{n}{FN_i}}
\label{eq:recall}
\end{equation}

where $TP_i$, $FP_i$, $FN_i$, with $i$ from 1 to $n$ (the number of media units in the dataset), represent the number of true positive, false positive and false negative annotations for media unit $i$. Finally, the micro-F1 score is computed as the harmonic mean of the micro-precision and micro-recall.

An important variable in the evaluation is the {\it media unit-annotation score threshold} for differentiating between a negative and a positive classification. Traditional crowdsourcing aims at reducing disagreement, and therefore corresponds to high values for this threshold. Lower values means accepting more disagreement in the classification of positive answers by the crowd. In our experiments, we tried a range of threshold values for each task, to investigate with which one we achieve the best results. The media unit-annotation score threshold was also used in gathering the set of trusted judgments for the evaluation (Section~\ref{subsec:expert_data}). All the data used in this paper can be found in our data repository\footnote{\url{https://github.com/CrowdTruth/Cross-Task-Majority-Vote-Eval}}.

\subsection{Trusted Judgments Collection}
\label{subsec:expert_data}

To perform the evaluation, a set of trusted judgments is necessary to assess the correctness of crowd annotations. For each dataset, we manually evaluated the correctness of all the media unit annotations that were generated by the crowd and the experts. Depending on the task, the number of media unit-annotation pairs can become quite high, so we explored methods to make the manual evaluation more efficient.

For the datasets that contain expert annotation, we calculated the thresholds which yielded the maximum agreement in number of annotations between the crowd and expert annotations.  These annotations were then added to the trusted judgments collection, as the judgment in this case is unambiguous.  The interesting cases appear when crowd and expert disagree.  Previous work we performed in crowdsourcing {\it Medical Relation Extraction}~\cite{aroyo2015truth} has indicated that experts might not always provide better annotations than crowd workers.  Additionally, for the {\it Sound Interpretation} task we noticed that experts provided considerably fewer tags than the crowd, and there was a large discrepancy between annotations of crowds and experts, with a very small overlap between their annotations.  Therefore, instead of simply relying on expert judgment, the annotations where crowd and expert disagree were manually relabeled by exactly one of the authors, and then added to the trusted judgments set, which is also published in our data repository. In Appendix~\ref{sec:appendix} we present a selection of examples where the expert judgment is different from the trusted judgment. While these cases might call into question the level of expertise of the domain experts, inconsistencies and disagreement in expert annotation are regularly reported in various annotation tasks~\cite{cheatham2014conference,mcdonnell2016relevant,INEL16.635}.  Furthermore, in Section~\ref{sec:results} we will show that using the trusted judgments for evaluation still results in the expert performing the best for 2 out of 3 tasks. The only task where the expert underperforms is {\it Sound Interpretation}, where the set of annotations provided by the expert is much smaller than the one provided by the crowd.

\begin{figure*}[!b]
\centering
\caption{CrowdTruth F1 scores for all crowdsourcing tasks.}
\label{fig:f1_mv}
\begin{subfigure}{.45\textwidth}
\includegraphics[width=\linewidth]{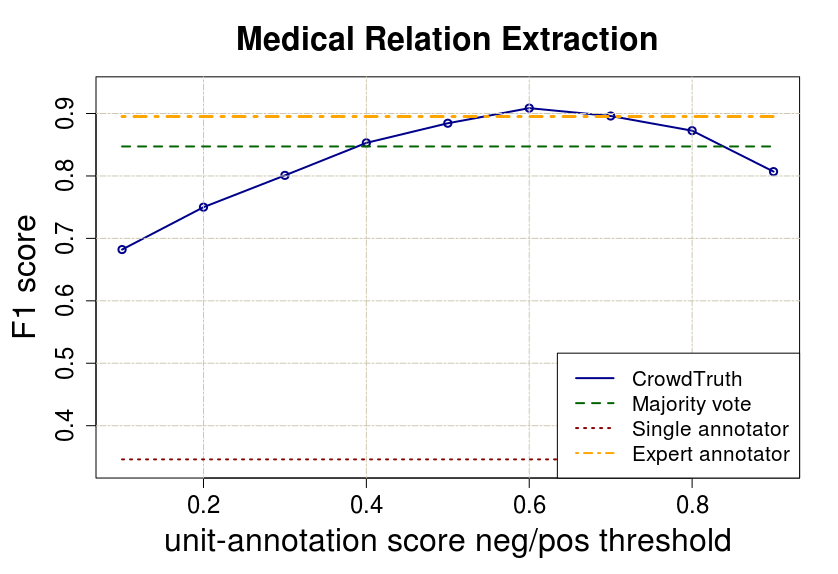}
\end{subfigure}%
\begin{subfigure}{.45\textwidth}
\includegraphics[width=\linewidth]{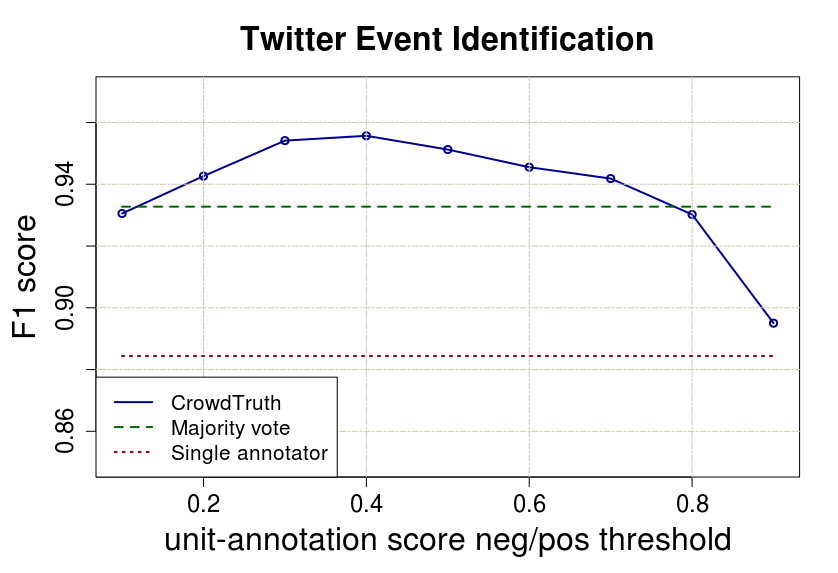}
\end{subfigure}
\begin{subfigure}{.45\textwidth}
\includegraphics[width=\linewidth]{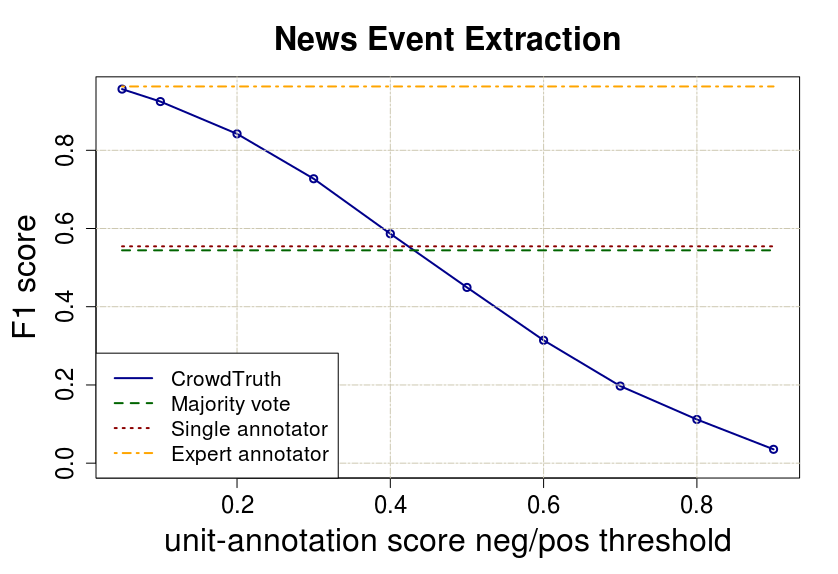}
\end{subfigure}%
\begin{subfigure}{.45\textwidth}
\includegraphics[width=\linewidth]{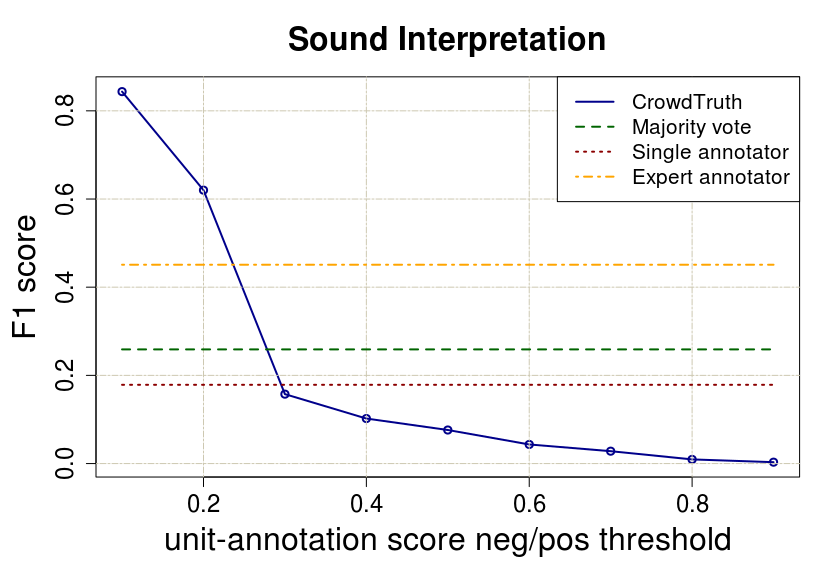}
\end{subfigure}
\end{figure*}

We collected expert annotations for the {\it Medical Relation Extraction} data by employing medical students. Each sentence was annotated by exactly one person. The annotation task consisted of deciding whether or not the UMLS seed relation discovered by distant supervision is present in the sentence for the two selected terms.

For the {\it Sound Interpretation} task, each sound in the dataset contains a description and a set of keywords that were provided by the authors of the sounds. We consider the keywords provided by the sounds' authors as trusted judgments given by domain experts.

The \emph{news event extraction} data was annotated with events by various linguistic experts. In total, 5 people annotated each sentence but we only have access to the final annotations, a consensus among the annotators. In the annotation guidelines described in \cite{pustejovsky2003timebank}, events are defined as situations that happen or occur, but are not generic situations. In contrast to the crowdsourcing task, where the workers had very loose instructions, the experts had very strict rules for identifying events, strictly based on linguistic features: \emph{(i)} tensed verbs: has called, will leave, was captured, \emph{(ii)} stative adjectives: sunken, stalled, on board and \emph{(iii)} event nominals: merger, Military Operation, Gulf War.

The only task without expert annotation is {\it Twitter Event Identification} -- as it is in the open domain, no experts exist for this type of data.

\section{Results}
\label{sec:results}

We begin by evaluating {\bf how the majority vote method compares with CrowdTruth}, by calculating the precision/recall metrics using the gold standards we collected for each of the four crowdsourcing tasks.  Figure~\ref{fig:f1_mv} shows the F1 score for CrowdTruth over the four tasks.  The results are calculated for different media unit-annotation score thresholds for separating the data points into positive and negative classifications.  Table~\ref{tab:f1_mv} shows the detailed scores for CrowdTruth, given the highest F1 media unit-annotation score threshold.

\begin{table*}[!tb]
\centering
\caption {CrowdTruth evaluation results, given the highest F1 media unit-annotation score threshold.}
\label{tab:f1_mv}
\begin{tabular}{|lr|ccccl|}
\hline
{\bf Task} & {\bf Dataset} & {\bf Precision} & {\bf Recall} & {\bf F1 score} & {\bf Accuracy} &  {\bf media unit-annotation score threshold} \\ \hline \hline

\multirow{4}{*}{\parbox{1.8cm}{{\bf Medical \\ Relation \\ Extraction}}} & CrowdTruth & 0.86 & 0.962 & 0.908 & 0.932 & 0.6 \\
 & expert & 0.899 & 0.89 & 0.895 & 0.927 &  \\
 & majority vote & 0.924 & 0.781 & 0.847 & 0.902 & \\
 & single & 0.222 & 0.776 & 0.346 & 0.748 & \\ \hline

\multirow{3}{*}{\parbox{1.8cm}{{\bf Twitter \\ Event \\ Identification}}} & CrowdTruth & 0.965 & 0.945 & 0.955 & 0.995 & 0.4 \\
 & majority vote & 0.984 & 0.885 & 0.932 & 0.984 & \\
 & single & 0.959 & 0.819 & 0.884 & 0.972 & \\ \hline

\multirow{4}{*}{\parbox{1.8cm}{{\bf News \\ Event \\ Extraction}}} & CrowdTruth & 0.984 & 0.929 & 0.956 & 0.931 & 0.05 \\
 & expert & 0.983 & 0.944 & 0.963 & 0.942 & \\
 & majority vote & 0.985 & 0.375 & 0.544 & 0.492 & \\
 & single & 0.99 & 0.384 & 0.554 & 0.501 & \\ \hline

\multirow{4}{*}{\parbox{1.8cm}{{\bf Sound \\ Interpretation}}} & CrowdTruth & 1 & 0.729 & 0.843 & 0.815 & 0.1 \\
 & expert & 1 & 0.291 & 0.45 & 0.515 & \\
 & majority vote & 1 & 0.148 & 0.258 & 0.418 & \\
 & single & 1 & 0.098 & 0.178 & 0.383 & \\ 
\hline
\end{tabular}
\end{table*}

\begin{table*}
\caption{$p$-values for McNemar's test of statistical significance in the CrowdTruth classification, compared with the others.}
\label{tab:stat_sig}
\centering
\begin{tabular}{|r|ccc|}
\hline
{\bf Task} & {\bf Maj. Vote} & {\bf Expert} & {\bf Single} \\ \hline \hline
Medical Relation Extraction & $0.0001$ & $0.629$ & $< 2.2 \times 10^{-16}$ \\ \hline
Twitter Event Identification & $0.0001$ & N/A &  $6.145 \times 10^{-15}$ \\ \hline
News Event Extraction &  $< 2.2 \times 10^{-16}$ & $0.505$ & $< 2.2 \times 10^{-16}$ \\ \hline
Sound Interpretation & $< 2.2 \times 10^{-16}$ & $< 2.2 \times 10^{-16}$ & $< 2.2 \times 10^{-16}$ \\ \hline
\end{tabular}
\end{table*}

Across all four tasks, the CrowdTruth method performs better than both majority vote and the single annotator dataset.  While majority vote unsurprisingly performs the best on precision, as a consequence of its lower rate of positive labels, CrowdTruth consistently scores the best for both recall, F1 score and accuracy.  These differences in classification are statistically significant, as shown in Table~\ref{tab:stat_sig} -- this was calculated using McNemar's test~\cite{mcnemar1947note} over paired nominal data.

The evaluation of CrowdTruth compared with the expert is more nuanced. For the {\it Medical Relation Extraction} and {\it news event extraction tasks}, CrowdTruth performs as well as the expert annotators, with p-values indicating there is no statistically significant difference in the classifications.  In contrast, for the task of {\it Sound Interpretation}, CrowdTruth performs better than the expert by a large margin.

\begin{figure*}[!tb]
\centering
\caption{The effect of the number of workers per unit on the F1 score, calculated at the best media unit-annotation score threshold (Table~\ref{tab:f1_mv}). For every point, the F1 is calculated with at most the given number of workers. The number of units used in the calculation of the F1 is shown in the y-axis on the right.}
\label{fig:f1_workers}
\centering
\begin{subfigure}{.45\textwidth}
\includegraphics[width=\linewidth]{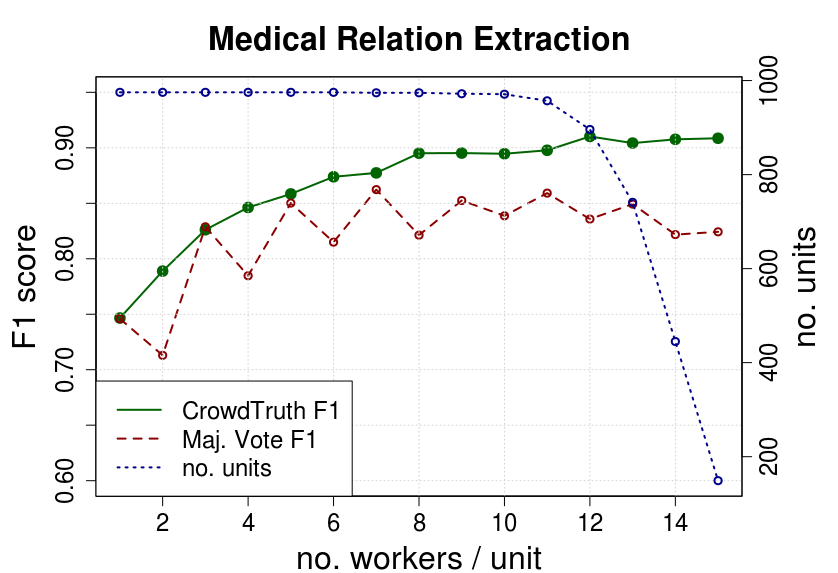}
\end{subfigure}%
\begin{subfigure}{.45\textwidth}
\includegraphics[width=\linewidth]{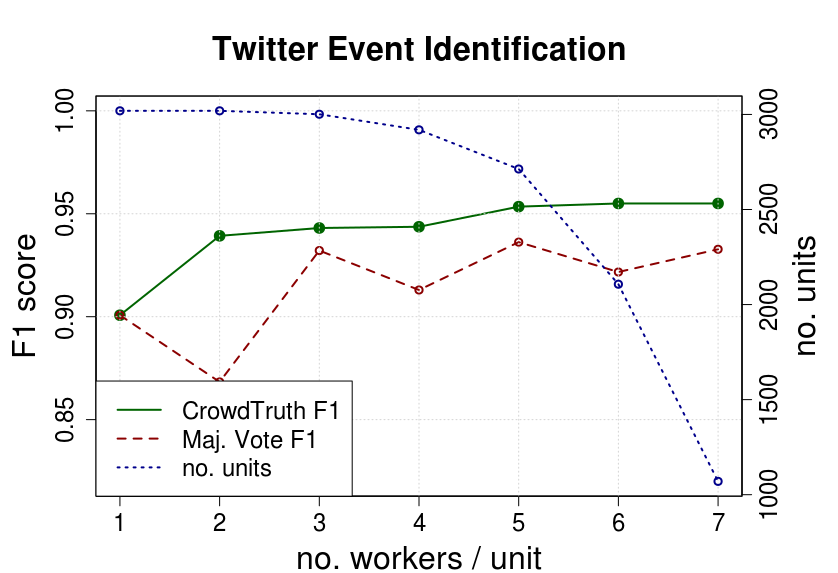}
\end{subfigure}
\begin{subfigure}{.45\textwidth}
\includegraphics[width=\linewidth]{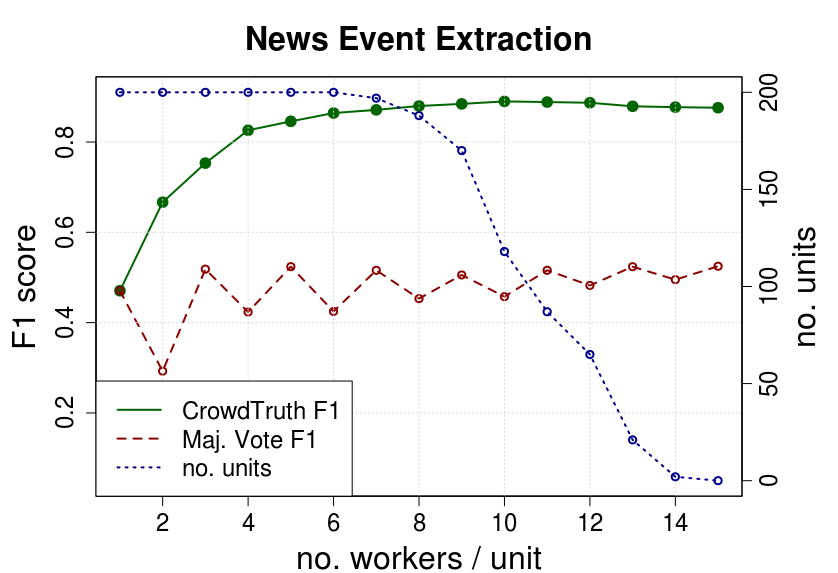}
\end{subfigure}%
\begin{subfigure}{.45\textwidth}
\includegraphics[width=\linewidth]{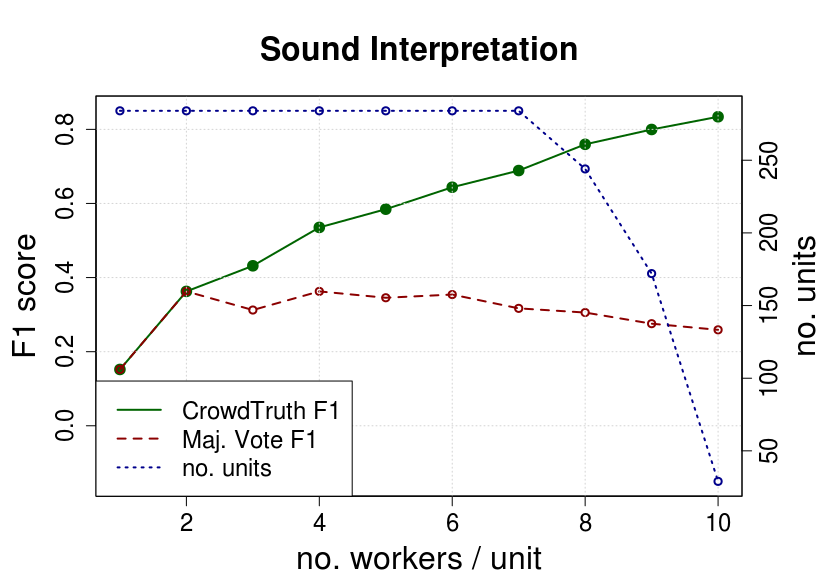}
\end{subfigure}
\end{figure*}

The second evaluation shows the {\bf influence of the number of workers on the quality of the CrowdTruth data}.  Figure~\ref{fig:f1_workers} shows the CrowdTruth F1 score in relation to the number of workers.  Given one task, the number of workers per unit varies because of spam removal, so the F1 score was calculated using at most the number of workers at every point in the graph. The number of units annotated with the given number of workers is also shown in the graph.

The effects of the number of workers on the CrowdTruth F1 is clear -- more workers invariably leads to a higher F1 score.  For the tasks of {\it Medical Relation Extraction}, {\it Twitter Event Identification} and {\it News Event Extraction}, the CrowdTruth F1 grows into a straight line, showing that the opinions of the crowd stabilize after enough workers.  For the {\it Sound Interpretation} task, the CrowdTruth F1 score is still on an upwards trend after 10 workers, possibly indicating that more workers are necessary to get the full spectrum of annotations.

Figure~\ref{fig:f1_workers} also shows that CrowdTruth performs better than majority vote regardless of the number of workers per task. For closed tasks, increasing the number of workers has a positive impact on the majority vote F1 score.  For open tasks, adding more workers has less of an effect -- more workers increase the size of the annotation set for a unit, which is typically larger than for closed tasks, but the agreement is low because opinions are split between possible annotations.

\begin{figure*}[!tbh]
\centering
\caption{CrowdTruth F1 score evaluation, using expert annotation as ground truth.}
\label{fig:f1_exp}
\begin{subfigure}{.45\textwidth}
\includegraphics[width=\linewidth]{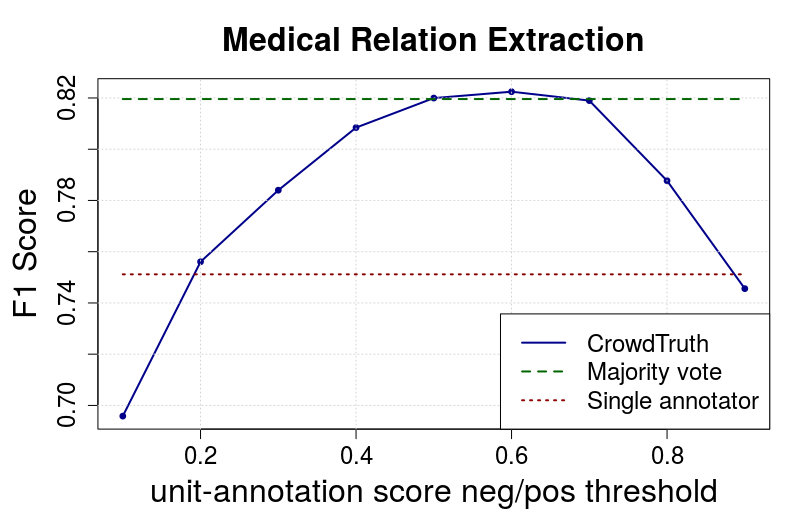}
\end{subfigure}%
\begin{subfigure}{.45\textwidth}
\end{subfigure}
\begin{subfigure}{.45\textwidth}
\includegraphics[width=\linewidth]{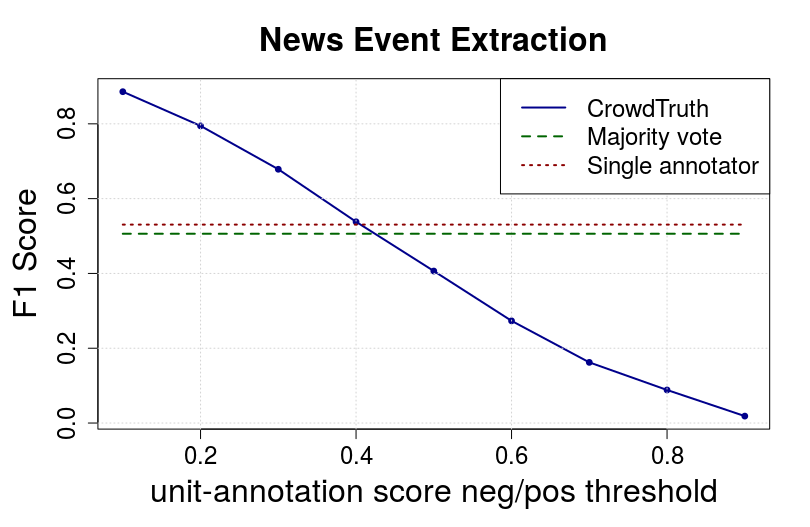}
\end{subfigure}%
\begin{subfigure}{.45\textwidth}
\includegraphics[width=\linewidth]{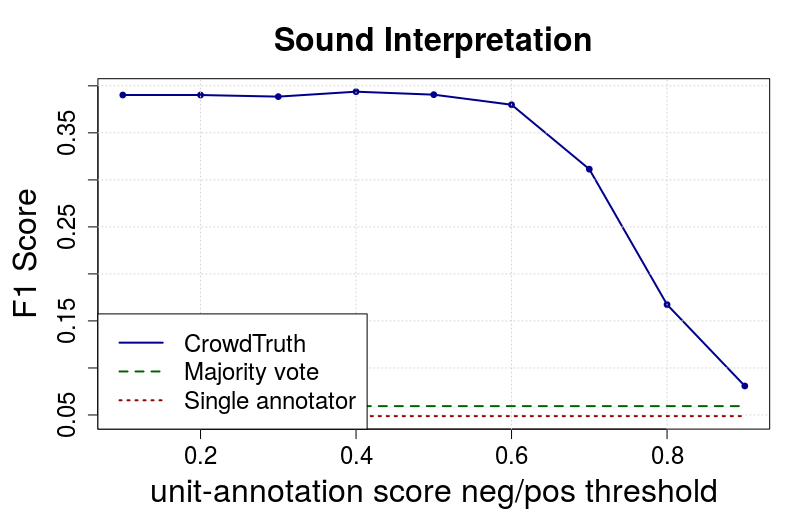}
\end{subfigure}
\end{figure*}

Finally, Figure~\ref{fig:f1_exp} shows an evaluation of CrowdTruth using only the expert annotations as ground truth (the {\it Twitter Event Identification} task does not have experts, so it could not be evaluated). The F1 scores are lower than in the evaluation over the trusted judgments collection. For the {\it Medical Relation Extraction Task}, majority vote performs essentially the same as CrowdTruth, whereas for the open-ended tasks, CrowdTruth still performs better. However, as we have shown in Appendix~\ref{sec:appendix}, the expert annotations contain errors and are sometimes incomplete, particularly in the case of open-ended tasks. The evaluation using expert ground truth was done to show that the trusted judgments set is not biased in favor of CrowdTruth.

\section{Discussion}
\label{sec:discussion}

The first goal in this paper was to show that the {\bf ambiguity-aware CrowdTruth approach with multiple annotators and disagreement-based quality scores can perform better than majority vote}, a method that enforces consensus among annotators.  Our results over several crowdsourcing tasks, as seen in Figure~\ref{fig:f1_mv}, show this clearly.

The gap in performance between CrowdTruth and majority vote is the most striking for open tasks ({\it News Event Extraction} and {\it Sound Interpretation}).  These tasks also require the lowest agreement threshold for achieving the best performance with CrowdTruth.  During the trusted judgments collection process, we observed how these tasks are prone to a wide range of opinions -- for instance, in the case of {\it Sound Interpretation}, there are frequent examples of labels that are semantically dissimilar, but could reasonably be applied to the same sound (e.g. the same sound was annotated with the tag {\tt balloon popping} by one worker, and with {\tt gunshot} by another worker).  Because of this, enforcing consensus does not work for these tasks, and ambiguity-aware annotation aggregation appeared to be a viable solution.

Our evaluation also shows that processing crowd data with ambiguity-aware metrics performs at least as well as expert annotators, which is not the case for majority vote.  Crowdsourcing annotation is significantly cheaper in cost than experts --  e.g. even with 15 workers per unit, crowdsourcing for the task of {\it Medical Relation Extraction} cost 2/3 of what the experts did. The crowd also has the advantage of being readily available on platforms such as Figure Eight, while the process of finding and hiring expert annotators can incur significant time costs. As our results showed, in order for the crowdsourcing to produce results comparable in quality to that of experts, appropriate processing with ambiguity-aware metrics is a necessity.

The variation in the optimal media unit-annotation score thresholds across the tasks shows that the level of ambiguity is dependent on the crowdsourcing task, thus supporting our triangle of disagreement model (Section~\ref{subsec:metrics}).  It is not surprising that the task with the highest agreement threshold ({\it Medical Relation Extraction}) also has the most exact definition of a correct answer (i.e. whether a medical relation is expressed or not in a given sentence).  The definition of a medical relation is fairly clear; in contrast, the definition of an event is more subjective, therefore workers were able to come up with a wider range of correct annotations.

The experimental setup provides an empirical method for selecting the optimal threshold for media unit-annotation score.  However, if performing an evaluation with trusted judgments is not possible, selecting the optimal threshold becomes more difficult.  For open-ended tasks, the experiments indicate that almost all opinions matter, and the agreement threshold should be as low as possible.  In these cases, spam workers can be successfully eliminated by in-task effort consistency checks, and there is no need to enforce agreement beyond that.  In contrast, the experiments for closed tasks show higher agreement thresholds tend to work better.  The difficulty as well as the subjectivity of the domain also appear to have an impact.  The threshold should grow together with the difficulty, and inversely with subjectivity.  However, both difficulty and subjectivity might be difficult to measure in practice.  In the end, the tuning of the threshold should be regarded similarly to a precision-recall trade-off analysis, where the optimal value depends on the requirements of the ground truth (high precision but many false negative crowd labels, or high recall but more false positives).  The high variability for optimal threshold values also shows the limitations of traditional evaluation metrics like precision and recall that rely on discrete labels.  CrowdTruth metrics were constructed to measure ambiguity on a continuous scale, but the use of standard metrics resulted in losing this information by forcing the conversion to either positive or negative. Ultimately, our goal is to move away from a binary ground truth that needs to be calculated using a fixed threshold, and instead to use the CrowdTruth metrics to express ambiguity on a continuous scale.

The second goal of the experiment was to show {\bf the effect of the number of workers on the quality of CrowdTruth annotations}.  The results in Figure~\ref{fig:f1_workers} clearly show the increase in F1 score for CrowdTruth as more workers contribute to the tasks. This combined with the poor performance of the single annotator dataset proves the importance in considering a large enough pool of workers to be able to accurately capture the full spectrum of opinions.

The stabilization of the F1 score for {\it Medical Relation Extraction}, {\it Twitter Event Identification} and {\it News Event Extraction} is an indication that we have indeed managed to collect the entire set of opinions for these tasks.  The fact that the scores all stabilize at different points in the graph (around 8 workers for {\it Medical Relation Extraction}, 5 for {\it Twitter Event Identification}, and 10 for {\it News Event Extraction}) indicates that the optimal number of workers is dependent on the task type, thus also confirming our hypothesis that more workers than what is typically being considered in crowdsourcing studies are necessary for acquiring a high quality ground truth.

There exists a trade-off between cost and quality of annotations that should also be considered when optimizing the number of workers.  The higher cost was justified for these tasks, as the expert annotation was three times more expensive than the crowdsourced annotations at expert quality level.

An interesting observation is that the optimal number of workers per task does not seem to influence the optimal media unit-annotation score threshold for the task.  The {\it News Event Extraction} requires a high number of workers, but the optimal media unit-annotation score threshold is low, while the {\it Twitter Event Identification} requires a low number of workers, and also a low media unit-annotation score threshold, at least compared to {\it Medical Relation Extraction}. 

While four tasks is a small sample to draw conclusions from, our findings seem to indicate that ambiguity in the crowdsourcing system has an impact on both the optimal number of workers per task, as well as the clarity of the media units.  These observations will form the basis for our future research in modeling crowd disagreement.

Finally, it is worth discussing the outlier characteristics of the {\it Sound Interpretation} task. It is the only task that does not achieve a stable F1 curve (Figure~\ref{fig:f1_workers}) possibly due to insufficient workers assigned to it. It is also unique in its lack of false positive examples -- precision is 1 for the optimal media unit-annotation score threshold (Table~\ref{tab:f1_mv}), meaning that all labels collected from the crowd were accepted as part of the trusted judgments, with the exception of the spam workers that were removed from the set.  {\it Sound Interpretation} is also the only task for which the expert annotator performed comparatively poor, with a statistically significant difference from CrowdTruth. As mentioned in the beginning of this section, after collecting the trusted judgments for this task, it became clear that the main challenge for the {\it Sound Interpretation} task is not to achieve consensus between annotators, but to collect the entire spectrum of annotations that describe a sound, given that this spectrum is so large (e.g. the tags {\tt balloon popping} and {\tt gunshot} can both reasonably apply to the same sound). For this reason, it was difficult to label tags as false positives, and the annotations of the workers, experts included, were largely non-overlapping, as they tended to interpret the sounds quite differently.  The {\it Sound Interpretation} task is therefore an extreme example of subjective ground truth.

\section{Related Work}
\label{sec:relatedwork}

\subsection{Crowdsourcing Ground Truth}

Crowdsourcing has grown into a viable alternative to expert ground truth collection, as crowdsourcing tends to be both cheaper and more readily available than domain experts. Experiments have been carried out in a variety of tasks and domains:  medical entity extraction~\cite{zhai2013web,Finin2010,van2012eu}, medical relation extraction~\cite{kilicoglu2011constructing,van2012eu}, open-domain relation extraction~\cite{kondreddi2014combining}, clustering and disambiguation~\cite{Lee2013}, ontology evaluation~\cite{noy2013mechanical}, web resource classification~\cite{castano2016human} and taxonomy creation~\cite{bragg2013crowdsourcing}. \cite{Snow2008} have shown that aggregating the answers of an increasing number of unskilled crowd workers with majority vote can lead to high quality NLP training data. The typical approach in these works is to assume the existence of a universal ground truth. Therefore, disagreement between annotators is considered an undesirable feature, and is usually discarded by using either of the following methods: restricting annotator guidelines, picking one answer that reflects some consensus usually through majority voting, or using a small number of annotators.

\subsection{Disagreement and Ambiguity in Crowdsourcing}

Besides CrowdTruth, there exists some research on how disagreement in crowdsourcing should be interpreted and handled. In assessing the OAEI benchmark, \cite{cheatham2014conference} found that disagreement between annotators (both crowd and expert) is an indicator for inherent uncertainty in the domain knowledge, and that current benchmarks in ontology alignment and evaluation are not designed to model this uncertainty. \cite{plank-hovy-sogaard:2014:P14-2} found similar results for the task of crowdsourced part-of-speech tagging -- most inter-annotator disagreement was indicative of debatable cases in linguistic theory, rather than faulty annotation. \cite{Bayerl2011} also investigate the role of inter-annotator disagreement as a possible indicator of ambiguity inherent in natural language. \cite{lau2014measuring} propose a method for crowdsourcing ambiguity in the grammatical correctness of text by giving workers the possibility to pick various degrees of correctness, but inter-annotator disagreement is not discussed as a factor in measuring this ambiguity. \cite{schaekermann2016} propose a framework for dealing with uncertainty in ground truth that acknowledges the notion of ambiguity, and uses disagreement in crowdsourcing for modeling this ambiguity. For the task of word sense disambiguation, \cite{jurgens2013embracing} show that, in modeling ambiguity, the crowd was able to achieve expert-level quality of annotations. \cite{Chang:2017:Revolt} implemented a workflow of tasks for collecting and correcting labels for text and images, and found that ambiguous cases cannot simply be resolved by better annotation guidelines or through worker quality control. Finally, \cite{lin2014re} shows that often, machine learning classifiers can achieve a higher accuracy when trained with noisy crowdsourcing data. To our knowledge, our paper presents the first experiment across several tasks and domains that explores ambiguity as a property of crowdsourcing systems, and how it can be interpreted to improve the quality of ground truth data.

\subsection{Crowdsourcing Aggregation beyond Majority Vote}

The literature on alternative crowdsourcing aggregation metrics typically focuses on analyzing worker performance -- identifying spam workers~\cite{Bozzon:2013,Kittur2008,Ipeirotis:2010}, and analyzing workers' performance for quality control and optimization of the crowdsourcing processes~\cite{Singer:2013}. \cite{NIPS2009_3644} and \cite{welinder2010multidimensional} have used a latent variable model for task difficulty, as well as latent variables to measure the skill of each annotator, to optimize crowdsourcing for image labels. \cite{werling2015job} use on-the-job learning with Bayesian decision theory to assign the most appropriate workers for each task, for both text and image annotation. Finally, \cite{prelec2017solution} show that the surprisingly popular crowd choice (i.e. the answer that most workers thought would not be picked by other workers, even though it is correct) gave better results than the majority vote for a variety of tasks with unambiguous ground truths (state capitals, trivia questions and price of artworks).

All of these approaches show promising improvements over the use of majority vote as an aggregating method.  These methods were developed only for closed tasks, primarily dealing with classification.  However, the novel approach of CrowdTruth allows to explore both closed and open-ended tasks.  Furthermore, our focus is on modeling ambiguity as a latent variable in the crowdsourcing system, as well as its role in generating inter-annotator disagreement, which these approaches currently do not take into account. We believe an optimal crowdsourcing approach would combine both ambiguity modeling, as well as specialized task assignment to workers. For instance, \cite{felt2015early} developed a generative model to aggregate crowd scores that incorporates features of the data (e.g. number of words), although they do not evaluate the performance of specific features. Ambiguity as measured with CrowdTruth, like the media unit-annotation score, could be used as a data feature in such a system.

\section{Conclusions}
\label{sec:conclusions}

Gathering human annotation is a major bottleneck in the process of knowledge base curation. Crowdsourcing-based approaches are gaining popularity in the attempt to solve the issues related to volume of data and lack of annotators. Typically these practices use inter-annotator agreement as a measure of quality. However, by ignoring inter-annotator disagreement, these practices tend to create artificial data that is neither general nor reflects the ambiguity inherent in the source.

In this paper we presented an empirically derived methodology for efficiently gathering of human annotation by aggregating crowdsourcing data with CrowdTruth metrics, which harness the inter-annotator disagreement. We applied this methodology over a set of diverse crowdsourcing tasks: closed tasks ({\it Medical Relation Extraction}, {\it Twitter Event Identification}), and open-ended tasks ({\it News Event Extraction} and {\it Sound Interpretation}).  Our results showed that the ambiguity-aware CrowdTruth approach allows us to collect richer data, which enables reasoning about the ambiguity of the content being annotated. This is intrinsically relevant to the Semantic Web community, i.e. to identify the semantics of ambiguity across all modalities, e.g. text, images, videos and sounds. Our results also showed that, in all the tasks we considered, such ambiguity-aware quality scores provide better ground truth data than the traditional majority vote. Moreover, we have shown that CrowdTruth annotations have at least the same quality, even better in the case of {\it Sound Interpretation}, as expert annotations.  Finally, we showed that, contrary to the common crowdsourcing practice of employing a small number of annotators, adding more crowd workers actually can lead to significantly better annotation quality.

In the future, we plan to expand our methodology to more complex annotation tasks, that require multiple or combined types of input beyond the closed/open-ended categorization we presented in this paper. We are also working on expanding the CrowdTruth metrics for ambiguity to incorporate the state-of-the art in modeling crowd worker and data features~\cite{felt2015early}. Finally, we want to use the CrowdTruth data in practice for training and evaluating information extraction models used to populate the Semantic Web.

\section*{Acknowledgements}

We would like to thank Emiel van Miltenburg for assisting with the exploration of feature analysis of sounds, Chang Wang and Anthony Levas for providing and assisting with the medical data, Zhaochun Ren for the help in gathering the Twitter dataset, Tommaso Caselli for providing the news dataset, and the anonymous crowd workers for their contributions to our crowdsourcing tasks.

% {\color{red} did you include a reference to the TIIS paper here too?}

\bibliography{sources}
\bibliographystyle{natbib}

\onecolumn

\appendix

\section{Example Media Units Where the Expert Judgment Is Different from the Trusted Judgment}
\label{sec:appendix}

\begin{table*}[!htb]
\centering
\caption {Example sentences from the {\it Medical Relation Extraction} task where the expert judgment is different from the trusted judgment. The pair of terms that express the medical relation are shown in italic font in the media unit.}
\label{tab:ex_relex}
%\resizebox{0.85\textwidth}{!}{
\begin{tabular}{|p{9cm}|c|ccc|}
\hline
{\bf Media Unit} & {\bf Annotation} & {\bf Expert} & {\bf Crowd} & {\bf Trusted} \\ 
 & & {\bf Judgment} & {\bf Score} & {\bf Judgment} \\ \hline \hline
The {\it epidermal nevus syndrome} is a neurocutaneous disorder characterized by {\it distinctive skin lesions} and often serious somatic and central nervous system (CNS) abnormalities. &	$cause$ & no & 0.98 & yes \\ \hline
For empiric $treat$ment of epididymitis, especially when gonococcal or {\it chlamydial infection} is likely Ofloxacin or {\it levofloxacin} should be used  only  if epididymitis is not $cause$d by gonorrhea. & $treat$ & no & 0.966 & yes \\ \hline 
In contrast, we did not find a definite increase in the LGL percentage within 6 months postpartum in patients with {\it Graves' disease} who relapsed into {\it Graves' thyrotoxicosis}. & $cause$ & no & 0.738 & yes \\ \hline 
The 1 placebo controlled trial that found black cohosh to be effective for {\it hot flashes} did not find {\it estrogen} to be effective, which casts doubt on the study's validity. & $treat$ & no & 0.73 & yes \\ \hline 
{\it Multicentric reticulohistiocytosis (MR)} is a {\it systemic disease} of unknown $cause$ characterized by the presence of a heavy macrophage infiltrate in skin and synovial tissues and the development of an erosive polyarthritis. & $cause$ & yes & 0.697 & no \\ \hline 
Urokise versus {\it tissue plasminogen activator} in {\it pulmonary embolism}. & $treat$ & yes & 0.365 & no \\ \hline 
The principal differences between these vaccines are the transmission of live vaccine viruses from recipients to their contacts and the occurrence of occasional cases of {\it paralytic poliomyelitis} associated with use of {\it live poliovirus vaccine} & $treat$ & yes & 0.1 & no \\ \hline 
These cases highlight the importance of considering {\it PTLD} in the differential diagnosis of {\it lymphadenopathy}. & $cause$ & yes & 0.09 & no \\ \hline 
\end{tabular}
%}
\end{table*}

\begin{table*}[!hb]
\centering
\caption {Example sentences from the {\it News Event Extraction} task where the expert judgment is different from the trusted judgment. The annotation is shown in italic font in the media unit.}
\label{tab:ex_news}
\begin{tabular}{|p{9cm}|c|ccc|}
\hline
{\bf Media Unit} & {\bf Annotation} & {\bf Expert} & {\bf Crowd} & {\bf Trusted} \\ 
 & & {\bf Judgment} & {\bf Score} & {\bf Judgment} \\ \hline \hline
The plan provides for the {\it distribution} of one common stock-purchase right as a dividend for each share of common outstanding & $distribution$ & no & 0.95 & yes \\ \hline 
Two Middle East terrorists with records of successful {\it attacks} against Western targets Abu Nidal and Abu Abbas have ties to Baghdad. & $attacks$ & no & 0.73 & yes \\ \hline 
Secretary of State James Baker said on ABC-TV's ``This Week With David Brinkley'' that the series of UN resolutions condemning Iraq's {\it invasion} of Kuwait ``imply that the restoration of peace and stability in the Gulf would be a heck of a lot easier if he and that leadership were not in power in Iraq.'' & $invasion$ & no & 0.53 & yes \\ \hline 
The company also said it continues to explore all options concerning the possible {\it sale} of National Aluminum's 54.5\% stake in an aluminum smelter in Hawesville Ky. & $sale$ & no & 0.24 & yes \\ \hline 
Yield on the issue was 7.88\% & $no$ $event$ & yes & 0.14 & no \\ \hline 
Har-Shefi said she heard Amir talk about killing Rabin but did not tell the police because she did not believe he was {\it serious}. & $serious$ & yes & 0 & no \\ \hline 
The American hope is that someone from within Iraq perhaps from the army 's professional ranks will step forward and push Saddam Hussein aside so that the country can begin recovering from the disaster. & $no$ $event$ & yes & 0 & no \\ \hline
\end{tabular}
\end{table*}

\begin{table*}[!t]
\centering
\caption {Example sounds from the {\it Sound Interpretation} task where the expert judgment is different from the trusted judgment.}
\label{tab:ex_sounds}
%\scalebox{0.8}{
\begin{tabular}{|p{7cm}|c|c|ccc|}
\hline
{\bf Media Unit URL} & {\bf Media Unit} & {\bf Annotation} & {\bf Expert} & {\bf Crowd} & {\bf Trusted} \\ 
 & {\bf Description} &  & {\bf Judgment} & {\bf Score} & {\bf Judgment} \\ \hline \hline

\multirow{3}{7cm}{\url{https://freesound.org/data/previews/21/21266_88803-hq.mp3}} & \multirow{3}{*}{jazz} & cymbals & no & 0.272 & yes \\
& & bangle & no & 0.136 & yes \\ 
& & rhythmic & no & 0.136 & yes \\ \hline

\multirow{3}{7cm}{\url{https://freesound.org/data/previews/26/26086_11477-hq.mp3}} & \multirow{3}{*}{chicken} & birds & no & 0.538 & yes \\
& & geese & no & 0.359 & yes \\ 
& & horns & no & 0.359 & yes \\ \hline

\multirow{3}{7cm}{\url{https://freesound.org/data/previews/35/35823_317782-hq.mp3}} & \multirow{3}{*}{weird drums} & music & no & 0.875 & yes \\
& & band & no & 0.145 & yes \\ 
& & disco & no & 0.145 & yes \\ \hline

\multirow{3}{7cm}{\url{https://freesound.org/data/previews/39/39329_404624-hq.mp3}} & \multirow{3}{*}{trip hop} & beat & no & 0.371 & yes \\
& & percussion & no & 0.371 & yes \\ 
& & chimes & no & 0.371 & yes \\ \hline

\multirow{3}{7cm}{\url{https://freesound.org/data/previews/41/41462_78779-hq.mp3}} & \multirow{3}{*}{beer glasses} & clicks & no & 0.242 & yes \\
& & clink & no & 0.242 & yes \\ 
& & ding & no & 0.242 & yes \\ \hline
\end{tabular}
%}
\end{table*}

\end{document}